\documentclass[aps,showpacs,nofootinbib]{revtex4}
\usepackage{epsf,latexsym}

\begin{document}

\title{Gravity from the entropy of light}
\author{Alessandro Pesci}
\email{pesci@bo.infn.it}
\affiliation
{INFN-Bologna, Via Irnerio 46, I-40126 Bologna, Italy}

\begin{abstract}
The holographic principle,
considered in a semiclassical setting,
is shown to have
direct consequences on physics at a fundamental level.
In particular, 
a certain relation is pointed out
to be {\it the} expression of holography
in basic thermodynamics.

It is argued moreover 
that through this relation
holography can be recognized
to induce gravity,
and an expression for the gravitational lensing is obtained
in terms of entropy over wavelength
of black-body radiation,
or, at a deeper level,
in terms of maximum entropy over associated space 
to the elementary bit
of information.
\end{abstract}

\pacs{04.60.-m, 05.70.-a, 03.67.-a}

\maketitle


$ $
\section{Introduction}
According to the holographic principle \cite{tHooft},
only a finite amount of information
is allowed to be stored in a region with given 
bounding area,
scaling the limit as the area itself.
Its string-theoretical realization was considered
in \cite{Susskind}.

The original motivation has been very general,
since the principle was introduced through combination
of gravitational collapse and 
the basic tenets of quantum mechanics. 
Somehow, the gravitational context is found to highlight
a fundamental redundancy,
not visible otherwise,
in the quantum-mechanical degrees of freedom 
used to describe the systems.
Through a semiclassical discussion,
the aim of this note is 
to take the reverse path
and to spot
consequences, if any, on basic physics,
once the holographic principle
is assumed as primeval starting point. 

The most general formulation of the holographic principle
at semiclassical level --i.e. with matter degrees of freedom living
in a continuous background spacetime--
is perhaps the generalized covariant entropy bound \cite{FMW},
which states that the matter entropy $S$ on a terminated lightsheet $L$
is bounded by (in Planck units, the units we will use
throughout this note)

\begin{eqnarray}\label{gceb}
S \leq \frac{\Delta A}{4},
\end{eqnarray}
where ${\Delta A}$ is the area difference
between the start- and end-surfaces of $L$,
and $S$ is calculated through the entropy current $s^a$
(assumed to exist),
$S = \int_{L} s^a \epsilon_{abcd}$, 
being $\epsilon_{abcd}$
the spacetime (which we take 4-dim)
4-volume form.
In this work, inequality (\ref{gceb}) is assumed to be the precise 
mathematical formulation of the holographic principle.   

We recall a condition for (\ref{gceb}) 
--or a consequence of it when (\ref{gceb}) is assumed as starting point--
we shall use throughout the paper,
which shows up
in circumstances in which the effects of spacetime curvature
are extremely tiny.
The condition, 
derived and discussed in \cite{Pesci2, Pesci3},
is as follows :
In a spacetime with Einstein's equation,  
inequality (\ref{gceb}) is
universally true
if and only if
a local (i.e. depending on the point)
lower-limiting spatial
scale $l^*$, unrelated to gravity, 
is assumed to exist in the description 
of statistical systems,
with

\begin{eqnarray}\label{lstar}
l^* \equiv \frac{1}{\pi} \frac{s}{\rho + p} =
\frac{1}{\pi T} \ \left( 1 - \frac{\mu n}{\rho + p} \right)
\end{eqnarray}
(where, for the last expression, use of Gibbs-Duhem relation is made). 
Here
$s$, $\rho$, $n$, $p$, $T$, $\mu$ are respectively
local entropy, mass-energy and number densities,
local pressure, temperature and chemical potential 
(having, this latter, any rest energy included). 
%

For thin plane layers of thickness $l$
--actually, that geometric configuration which to the utmost
challenges the bound-- 
this becomes

\begin{eqnarray}\label{condition1}
l^* \leq 
l,
\end{eqnarray}
meaning that below $l^*$ the notions themselves of energy, entropy and
pressure in the layer become somehow undefined, or, equivalently,
that layers thinner than $l^*$ cannot be cutted physically if the assigned
values of the thermodynamic parameters have to remain unchanged after
cutting.
%
The reason why this limiting scale $l^*$ should exist
can be recognized through consideration 
of the trivial lightsheets associated with thin plane layers;
for them,
$\Delta A$  in (\ref{gceb})
is, from Raychaudhuri's equation \cite{Wald_book1}, 
quadratic in $l$, whereas $S$ is (obviously) linear,
so that, if a lower limit would not be envisaged for $l$, 
when $l \rightarrow 0$ inequality (\ref{gceb}) would be definitely violated.
%
The bound (\ref{gceb}) 
is attained
iff i) we consider plane layers
and ii) their thickness just attains the bound (\ref{condition1}).
$l^*$, considered as a time, instead of space, lower-limiting scale
(i.e. a lower-limiting time scale in the evolution
of statistical systems given by
the time it takes light to travel a distance $l^*$),
leads also to foresee \cite{Pesci4}
the universal bound 
to the relaxation times \cite{Hod}
of perturbed thermodynamic systems. 

As discussed in \cite{Pesci5},
from a quantum-mechanical standpoint
relation (\ref{condition1}) can be re-expressed as

\begin{eqnarray}\label{condition2}
l^* \leq
\lambda,
\end{eqnarray}
where $\lambda$ is the `typical'
quantum wavelength of constituent particles;
a notion which will be sharpened later.
What is meant here is that,
assuming that quantum mechanics allows 
for unaltered thermodynamic potentials
in physically cutted slices as thin as the $\lambda$ itself
of the constituent particles,
i) $\lambda$ is the minimum $l$ quantum-mechanically allowed
and ii) (\ref{condition1}) leads to (\ref{condition2}).
For a given system,
$\lambda$ in (\ref{condition2}) 
provides the tightest bound coming from holography 
to that combination
of thermodynamic potentials which we denote with $l^*$.
Only material media with $l^* = \lambda$
can attain the bound (\ref{gceb}), which is indeed attained
when (lightsheets trivially constructed on)
plane layers are considered with thickness $l = \lambda$. 

\section{A holographic law in basic thermodynamics}
In condition (\ref{condition2}) any connection with gravity, or
curvature, has disappeared.
We are left with
a flat-spacetime condition,
which compares
a combination of the thermodynamic potentials of a system 
with the quantum size of constituent particles; 
still, this condition is
an implication of or a pre-requisite for holography.
The aim of this Section
is just to emphasize that, due to this, 
(\ref{condition2}) 
can be read 
as a sort of
basic law of thermodynamics of holographic origin,
and discuss its nature.

Let us look, first, 
at that
a well-known basic thermodynamic bound
connecting energy $E$, entropy $S$ and size
--the circumscribing radius $R$ actually-- 
of a system
has already been long since proposed.
The Bekenstein universal bound to specific entropy \cite{BekSE}

\begin{eqnarray}\label{ueb}
\frac{S}{E} \leq 2 \pi R
\end{eqnarray}
indeed,
though 
originally found through an argument involving black hole physics,
was since the beginning recognized as 
a fundamental thermodynamic bound having nothing to do
with gravity.

In \cite{Pesci5} the relation of bound (\ref{ueb})
with holography has been discussed
(see also \cite{Boussobek}).
At first sight the bounds (\ref{ueb}) and 
the holographic relation (\ref{condition1})
seem quite different.
One difference is that 
in (\ref{condition1}) 
$p$ appears also (in $l^*$). 

As this regard
we point out that, 
considering the conditions of the original argument
bringing to (\ref{ueb}) through the use of the
generalized second law \cite{Bekgen1,Bekgen2},
in circumstances
more general than those originally considered
a contribution from 
the work done by pressure should also
be present.
 
The basic fact used in the derivation of (\ref{ueb})
is, indeed, that if a body with energy $E$
and circumscribing radius $R$ (and negligible self-gravity)
is swallowed by a black hole,
a lower limit definitely exists to the increase
of surface area of the hole, given by $8 \pi E R$
\cite{Bekgen1,Bekgen2}.
Then, from this, and imagining that given a body 
a process can always be found
for which this limit is attained,
through the use of the generalized second law for such a process
(\ref{ueb}) is obtained.

Now,
if we consider for simplicity a static black hole 
and, instead of assuming that a whole body is swallowed,
we dump in,
just when it
is at its first contact with the horizon,
a small element
of proper thickness $l$ and
cross-sectional area $A$
of an indefinitely extended fluid
(i.e. if, contrary to \cite{Bekgen1,Bekgen2}, 
we no longer require the stress tensor $T_{ab}$ to be vanishing
outside the body), for a perfect fluid 
assumed momentarily at rest in the local static frame
of the metric 
to first order in $T_{ab}$ 
(cf. \cite{Wald_book2})
we get
\begin{eqnarray}\label{deltam}
\Delta M =
\int_0^l dl^\prime \int d^2 S \ T_{ab} \xi^a k^b =
\kappa \int_0^l dl^\prime A l^\prime (\rho + p) =
\kappa A (\rho + p) \frac{l^2}{2} =
\kappa \frac{l}{2} (E + pV),
\end{eqnarray}
where $V$ is the proper volume of the fluid element,
$\kappa$ is the surface gravity of the hole,
$l^\prime$, the proper length in the fluid local frame,
is the chosen affine parameter for the null geodesics
(with tangent $k^a$) on the horizon,
$d^2 S$ is the cross-section element of the horizon at $l^\prime$ 
and $\xi^a$ is the Killing field, orthogonal to the horizon, 
which is timelike at infinity.
%
In (\ref{deltam}), 
use of the relation $\kappa l^\prime k^a = \xi^a$
with $\kappa = $ const
is made, which is appropriate in so far as
the change in the black hole geometry in the process 
can be neglected,
which always is the case
provided we choose $M$ large enough.
Here we are also assuming $V$ small enough
to allow for thermodynamic potentials approximately
constant in it.

From (\ref{deltam}) we get
the minimum horizon area increase, 
which is
$\frac{8\pi}{\kappa} \Delta M = 4\pi l (E + p V)$,
from which,
through use of the generalized second law,
we obtain

\begin{eqnarray}\label{mod_ueb}
\frac{S}{E + p V} \leq \pi l.
\end{eqnarray}
So,
if the element we drop in is, say,
a gas contained in a box,
and if the thermodynamic system we are considering
consists of both the gas {\it and} the constraining walls,
we are led to $S/E \leq \pi l$
(as in the derivation in \cite{Bekgen2}),\footnote{We are grateful
to R. Bousso for correspondence on this point.}
if instead our system
consists of the gas alone, 
our argument says that
the fundamental bound
should be given
by expression (\ref{mod_ueb}), 
i.e. with the term $pV$.

Inequality (\ref{mod_ueb}) manifestly coincides 
with condition (\ref{condition1}). 
In \cite{Pesci5},
it has been shown that, 
even if we start from (\ref{mod_ueb}), 
for macroscopic bodies we are lead anyway to (\ref{ueb})
(for whichever strength, indeed, of the gravitational effects).

Still,
the bound expressed by condition (\ref{condition2})
strongly differs from the Bekenstein bound, 
being the former actually enormously tighter than the latter.
%
For
a spherical homogeneous system with radius $R$, for example,
the Bekenstein bound says that the ratio $S/E$ is bounded by something
orders of $R$, while according to (\ref{condition2}) this same ratio
is bounded by orders of $\lambda$, the `typical' wavelength of
constituent particles, with 
always $\lambda \leq R$, 
and 
in general $\lambda \ll R$. 
Thus,
holography, which, too, has among its consequences the Bekenstein bound,
can be seen to imply for basic thermodynamics
a bound, condition (\ref{condition2}),
in general extremely stronger than Bekenstein's one. 
Bound (\ref{condition2}) 
seems thus could be considered as the fundamental requirement
in basic thermodynamics 
of the whole gravity-thermodynamics connection.
It is the basic-thermodynamic imprint 
of holography.

Let us give a closer look to bound (\ref{condition2}).
In the transition
from (\ref{condition1}) 
to (\ref{condition2}),
the particle `typical' wavelength $\lambda$
is, we said, a sort of minimum thickness
below which, in view of the uncertainty relations,
the value of thermodynamic potentials in a physically-cutted slice
are found different than before cutting.
It can be defined more precisely 
considering that a
limiting thickness $l_{min}$ should exist
for which the quantum spread in momentum $\Delta p_x^{ind}$, induced
by constraining the particles
in the given thickness,
becomes equal to
the intrinsic momentum spread $\Delta p_x^{int}$
(dictated by the assigned thermodynamic conditions)

\begin{eqnarray}\label{spread}
\Delta p_x^{ind} = 
\Delta p_x^{int}
\end{eqnarray} 
(here $x$ labels the direction orthogonal to the slice).
For constituent particles
all with a same 
quantum spatial uncertainty
which still we denote $\lambda$,
condition (\ref{spread}) 
will be reached 
by definition 
just when $l_{min} = \lambda$.
In the general case we define $\lambda$
as that thickness which gives (\ref{spread}).
For a Boltzmann  gas the value of a so-defined $\lambda$
is close to thermal de Broglie wavelength.

When checked on actual systems,
condition (\ref{condition2}) is found 
in general satisfied by far \cite{Pesci3}.
%
For the most entropic systems
it appears, instead, 
practically attained.
For black body radiation
--and in general for ultrarelativistic constituent particles--,
for example,
an argument described in \cite{Pesci3}
suggests 

\begin{eqnarray}\label{blackbody}
\lambda = 1/\pi T,
\end{eqnarray}
and thus 
$l^* = 
\frac{1}{\pi T} \ \left( 1 - \frac{\mu n}{\rho + p} \right) = 
\lambda$,
being $\mu = 0$. 
This prompts to consider
the uncertainty relations
as the mechanism which leads
to (\ref{condition2}), i.e.
to (\ref{condition1}).

From this perspective,
condition (\ref{condition2}) (and, thus, (\ref{condition1}))
appears to be a basic thermodynamic relation
arising from quantum mechanics alone.
%
This relation,
when combined with Einstein's lensing,
leads inexorably to the generalized
covariant bound to entropy,
that is to holography,
and vice versa is
the unavoidable consequence
of the latter in basic thermodynamics. 

As known,
the generalized second law
follows from the generalized entropy bound,
assuming the validity of the ordinary second law \cite{FMW}.
From the above this means that,
besides ordinary second law,
just another ingredient from basic
thermodynamics is 
needed, and is enough,
to give in a gravitational context
the generalized second law: relation (\ref{condition2}).
We see that this relation turns out to be
that component of basic thermodynamics
responsible for the `generalized part'
of the generalized second law,
i.e. that part which deals with processes
involving horizons.

If we imagine to start from the holographic principle
without any notion of quantum mechanics,
bound (\ref{gceb}) (which, consistently, should thus be understood as 
the log  
of the number of allowed different microscopic configurations
(instead of the log of the number of allowed orthogonal quantum states))
requires, we have seen, 
the existence of a lower-limiting scale $l^*$.
The existence of this spatial limit 
would suggest that a `size'  should be assigned
to the elementary constituents of matter,
while the value of the spatial limit (expression (\ref{lstar}))
would imply that,
at least for ultrarelativistic systems,
this `size' of the constituents should be related to their momentum
by the uncertainty relations.
That is,
holography demands for
a microscopic description of matter,
discrete in itself anyhow
if finite values are to be assigned to the entropy
of generic systems,
which too is driven 
by what we know as
the uncertainty relations,
i.e. it somehow demands for quantum mechanics.\footnote{Cf. \cite{Boussoqm};
see also \cite{Pesci4, Pesci5}.}

\section{A statistical origin for gravity}
At the end, what we have seen so far is that,
if the bundles of light rays actually shrink gravitationally,
the lower-limiting length $l^*$,
which manifests quantum mechanics,
is the effect of bound (\ref{gceb}).
%
But, what about the source of this shrinking ?
Looking at (\ref{gceb}),
we see that holography
demands that,
even in absence of interactions of any kind,
some mechanism must be at work
which shrinks the bundles of light rays
when going through matter.
The mere existence of some entropy
in a region requires, there, a focusing.
Bound (\ref{gceb}) moreover knows nothing else than 
entropy and focusing, so that it is quite natural,
as far as we take (\ref{gceb}) as our primeval starting point,
to somehow suspect entropy as the source of focusing.

Massive bodies can be considered
for which the entropy, when temperature is near absolute zero,
can be negligibly small, 
but the focusing sizeable.
So, if entropy is to be responsible of focusing, which entropy
should we take ?
%
It is clear that if,
for assigned $\rho$, $p$ and $\lambda$,
the rate of shrinking could be determined by a value of $s$ as high
as the limit $\frac{s}{\rho + p} = \pi \lambda$ in (\ref{condition2}),
any matter would comply with the bound (\ref{gceb}).
Thus,
the rate of shrinking could be set by the request that
bound (\ref{gceb}) be always satisfied,
and exactly attained by the most entropic systems. 

The perspective we advocate here is that
in giving some piece of matter what we are really doing
is to allocate some maximum amount of information or entropy,
let us call it `intrinsic' information/entropy,
associated with it.
%
The bound then says that this `intrinsic' 
information/entropy must focus light rays
at a precise rate.

We can derive the value of this rate
through consideration of material systems
which do attain the limit in (\ref{condition2}). 
For the terminated lightsheet
of a plane layer of a photon gas
with thickness $l$ as small as the limit
$l = l_{min} = \lambda = 1/\pi T$,
entropy just attains the limit in (\ref{gceb}),
as well as in (\ref{condition1}) and (\ref{condition2}).  
This means that at these conditions the shrinking
of the null congruence traversing the layer
is given by

\begin{eqnarray}\label{shrinking}
-\Delta A = 
4 S =
4 s A \lambda =
4 \pi (\rho + p) A \lambda^2 =
4 \pi A \lambda^2 T_{ab} k^a k^b,
\end{eqnarray}
where $T_{ab}$ is the stress energy of the photon gas
and $k^a$ are the tangents to the null congruence
with respect to the parametrization given by $l$,
and use of (\ref{condition2})
has been made.
On the other hand, from geometry
the shrinking is connected,
through Raychaudhuri's equation, 
to the Ricci tensor $R_{ab}$.
%
The Raychaudhuri equation, in our circumstances of
vanishing shear
and initially vanishing expansion $\theta = \frac{1}{A} \frac{dA}{dl}$,
for very small $l$ reads

\begin{eqnarray}\label{ray}
\frac{d\theta}{dl} =
-R_{ab} k^a k^b =
\text {const},
\end{eqnarray}
from which we get
$\theta = -l R_{ab} k^a k^b$
and thus the mentioned quadratic dependence
of $\Delta A$ on $l$ has the form

\begin{eqnarray}\label{geometry}
-\Delta A =
-\int_0^l \theta A dl^\prime =
A \frac{l^2}{2} R_{ab} k^a k^b.
\end{eqnarray}
%
From equation (\ref{shrinking}) this gives

\begin{eqnarray}\label{lensing}
R_{ab} k^a k^b = 8 \pi T_{ab} k^a k^b.
\end{eqnarray}

This is not a surprise.
%
The lensing turns out to be 
just that given by Einstein's equation. As it must be,
since bound (\ref{gceb}) has 
Einstein's lensing built-in.
%
The real point here is the perspective: the focusing
is determined by (`intrinsic') information/entropy 
through holography (bound (\ref{gceb})).
That is to say,
starting from holography without any notion of gravity
and knowing only of information/entropy, 
we end up with what we call gravity,
and this points
to a direction akin to 
\cite{Padma1, Padma2}
(last Section of both) and
\cite{Verlinde},
to some extent.\footnote{We notice
that in \cite{LeeKimLee} another derivation of gravity is given,
independent, like the present attempt,
of \cite{Padma1, Padma2, Verlinde}.
In it, gravity is derived
from the Laundauer principle
in quantum information theory as applied to horizons.
In \cite{CaravelliModesto}, moreover, 
the Einstein-Hilbert action,
or more general actions, are derived   
starting from black hole entropy
as calculated within loop quantum gravity.}

In \cite{Jacobson},
a thermodynamic interpretation of Einstein's equation
(in which this reveals itself as an equation of state) 
has already been given from an assumed
proportionality of horizon entropy and area.
The present attempt is supposed to provide a step forward
in that, 
through consideration of information as fundamental, 
the existence and strength of what we call gravity 
is reduced to a principle of 
maximum allowed amount of information 
inside any closed surface.
That is, what really matters here
is not just the thermodynamic nature of Einstein's equation
(a point, this, of paramount importance indeed, 
as for its implications on the opportunity of any attempt to quantize 
this equation as well as for its accounting of the occurrence
of thermodynamic laws for classical black holes \cite{Jacobson}),
but that the occurrence itself of gravity
is understood as
what must happen in order that
a certain primeval property of entropy be preserved.

Looking at (\ref{shrinking}), 
we see that 
the role played by $\pi(\rho + p)$
in determining the shrinking
of the congruence by the photon gas,
can be viewed as played by $s/\lambda$.
%
This suggests that, given some matter,
it is the `intrinsic' entropy density 
over wavelength,
namely the maximum entropy density over wavelength
at the given energy density+pressure,
what should be considered
as the proper source of focusing,
and what determines
the gravitational acceleration.

Considering light rays traversing orthogonally 
a thin plane layer of matter,
using the focusing equation \cite{MTW},
which, since the shear is vanishing, reads

\begin{eqnarray}\label{focusing}
\frac{d^2 A^{1/2}}{dl^2} =
- \frac{1}{2} R_{ab} k^a k^b A^{1/2},
\end{eqnarray}
we have that, assuming rotational symmetry
around the propagation axis (as it is the case if local matter
is assumed to be the only source of the field),
the local-frame acceleration $a_t$ felt by the photons of the
congruence 
while going through matter
being a distance $d$ apart can be expressed as

\begin{eqnarray}\label{tidal}
a_t = 
- \frac{1}{2} R_{ab} k^a k^b d =
- 4 \frac{s}{\lambda} d,
\end{eqnarray}
denoting with $s/\lambda$ the `intrinsic' entropy density 
per unit wavelength.
This same expression can be used to determine also 
the local acceleration with respect to the origin,
taken halfway between the photons, if $d$ changes its meaning
becoming the
distance from the origin. 

Expression (\ref{tidal}) fixes the gravitational
acceleration felt by photons in the local frame, 
due to the presence
of local matter,
as determined by its `intrinsic' entropy.
%
The operational meaning is that
given a material medium
with some local values $\rho_m$ and $p_m$ of energy density 
and pressure,
the local matter affects through holography the motion of photons
in the way expressed by (\ref{tidal})
(and this effect is what we call gravity),
where $s/\lambda$ is the entropy density
per unit wavelength
of a photon gas having energy density 
$\rho = \frac{3}{4} (\rho_m + p_m)$
and $\lambda = 1/\pi T$, where $T$
is its black body temperature.
The acceleration in (\ref{tidal}) is thus expressed
in terms of entropy density per unit wavelength of
that black body radiation which gives 
the `intrinsic' informational content of the local matter
we are considering.

\section{Concluding remarks}
In conclusion, what we have tried to show in the paper is that
the holographic principle, 
assumed as primeval starting point,
implies
both a basic relation in flat-spacetime thermodynamics,
relation (\ref{condition2})
(argued to be more fundamental than the Bekenstein bound),
and the curvature effects we call gravity,
with a new entropy 
--different from actual thermodynamic entropy--,
the `intrinsic' entropy (per $\lambda$) of a body, playing the role
of source of the curvature.
An expression of the gravitational acceleration in terms
of it has also been given (relation (\ref{tidal})). 

We can summarize what we have seen as follows.
The Einstein's focusing we have obtained,
or the explicit expression 
for the gravitational acceleration in equation (\ref{tidal}), 
permits to view
what we call `gravitational effects'
as actually holography at work,
and the `gravitational' acceleration as a `holographic' acceleration.
Gravity is merely all what is needed
for the `holographic' property of entropy to be preserved.
In particular in equation (\ref{tidal}) 
we can read directly the strength 
of a gravitational acceleration
per given amount of `intrinsic' information associated with matter. 

Relation (\ref{condition2})
(with expression (\ref{lstar})),
in the form of uncertainty-like relations,
establishes the rule
for finding the `intrinsic' information allocated with the assigned matter
(so that the meaning of the uncertainty relations would be
in their being what provides the informational content of matter).
This information (per $\lambda$)  is defined as
the maximum entropy per $\lambda$ we can associate to that matter,
i.e. the value which just attains (\ref{condition2});
and it turns out to be the entropy per $\lambda$ of 
`equivalent' (in a definite sense) black body radiation. 
This choice is dictated by the request
that
the strength of the holographic focusing 
be just that needed for the entropy in a (terminated) lightsheet
to be universally bounded 
(i.e. for every lightsheet geometry)
by the number $\Delta A/4$,
and just attained for the most challenging geometric choices.
Thus, holography,
by saying that the number of allowed
degrees of freedom
inside a given closed surface is bounded 
(by a value proportional to the area of the boundary),
induces
curvature effects determined by the `intrinsic' degrees of freedom 
carried by matter
(effects with a strength depending on the value of the bound), 
and this constitutes 
what we call gravity.

To speak of entropy per wavelength of black body radiation
means to speak of entropy per 1-bit-of-information thickness,
since the single bits of information are carried
by the single constituent photons
and we have $\simeq 1$ photon every $\lambda^3$ volume. 
This suggests a deeper description of what
we have discussed.
Indeed, holography can be stated to imply that
the allowed number of elementary bits of information
in a layer of 1 bit thickness
at given sum of energy and pressure energy in the layer 
is bounded.
To the extent that concepts like bit of information
and energy and pressure energy of a bit
can be regarded as primeval and, as such,
meaningful even in absence of space,
holography is pre-existing to space 
(cf. \cite{Padma1, Padma2, Verlinde},
and \cite{MarSmolin}).
In this perspective,
when space is introduced as the information on
`where' information is, 
the energy in the bit should spread
to keep unchanged
the elementary amount of information for the bit,
and this would be quantum mechanics.
When expressed in terms of this notion of
space, holography would then become the metric theory
which describes gravity.

I am grateful to Alessio Orlandi for fruitful discussions
on some of the arguments considered in the note.


\begin{thebibliography}{00}
\bibitem{tHooft} G. 't Hooft,
``Dimensional reduction in quantum gravity'',
essay dedicated to Abdus Salam,
published in {\it Salamfest} (1993) 0284,
gr-qc/9310026.

\bibitem{Susskind} L. Susskind,
``The world as a hologram'',
J. Math. Phys. {\bf 36} (1995) 6377,
hep-th/9409089.

\bibitem{FMW} $\acute {\rm E}$.$\acute {\rm E}$. 
Flanagan, D. Marolf and R.M. Wald,
``Proof of classical versions of the Bousso entropy bound and of the 
generalized second law'',
Phys. Rev. D {\bf 62} (2000) 084035, 
hep-th/9908070.

\bibitem{Pesci2} A. Pesci,
``From Unruh temperature to the generalized Bousso bound'',
Class. Quantum Grav. {\bf 24} (2007) 6219, 
arXiv:0708.3729.

\bibitem{Pesci3} A. Pesci,
``On the statistical-mechanical meaning of the Bousso bound'',
Class. Quantum Grav. {\bf 25} (2008) 125005,
arXiv:0803.2642.

\bibitem{Wald_book1} R.M. Wald,
{\it General Relativity} 
(The University of Chicago Press, Chicago, 1984).

\bibitem{Pesci4} A. Pesci, 
``A note on the connection between the universal relaxation bound and the 
covariant entropy bound'',
Int. J. Mod. Phys. D {\bf 18} (2009) 831,
arXiv:0807.0300.

\bibitem{Hod} S. Hod,
``Universal bound on dynamical relaxation times and black-hole quasinormal  
ringing'',
Phys. Rev. D {\bf 75} (2007) 064013,
gr-qc/0611004.

\bibitem{Pesci5} A. Pesci, 
``A proof of the Bekenstein bound for any strength of gravity 
through holography'',
Class. Quantum Grav. {\bf 27} (2010) 165006,
arXiv:0903.0319.
 
\bibitem{BekSE} J.D. Bekenstein,
``Universal upper bound on the entropy-to-energy ratio  
for bounded systems'',
Phys. Rev. D {\bf 23} (1981) 287.

\bibitem{Boussobek} R. Bousso,
``Light-sheets and Bekenstein's bound'', 
Phys. Rev. Lett. {\bf 90} (2003) 121302,
hep-th/0210295.

\bibitem{Bekgen1} J.D. Bekenstein,
``Black holes and entropy'',
Phys. Rev. D {\bf 7} (1973) 2333.

\bibitem{Bekgen2} J.D. Bekenstein,
``Generalized second law of thermodynamics in black-hole physics'',
Phys. Rev. D {\bf 9} (1974) 3292.

\bibitem{Wald_book2} R.M. Wald,
{\it Quantum field theory in curved spacetime and black hole thermodynamics} 
(The University of Chicago Press, Chicago, 1994).

\bibitem{Boussoqm} R. Bousso,
``Flat space physics from holography'',
JHEP{\bf 05}(2004) 050,
hep-th/0402058.

\bibitem{Padma1}
T. Padmanabhan,
``Thermodynamical Aspects of Gravity: New insights'',
Rept. Prog. Phys. {\bf 73} (2010) 046901,
arXiv:0911.5004.

\bibitem{Padma2}
T. Padmanabhan,
``Equipartition of energy in the horizon degrees of freedom and the emergence of gravity'',
Mod. Phys. Lett. A {\bf 25} (2010) 1129,
arXiv:0912.3165.

\bibitem{Verlinde} E.P. Verlinde,
``On the origin of gravity and the laws of Newton'',
arXiv:1001.0785.

\bibitem{LeeKimLee} J.-W. Lee, H.-C. Kim and J. Lee,
``Gravity from quantum information'',
arXiv:1001.5445.

\bibitem{CaravelliModesto} F. Caravelli and L. Modesto,
``Holographic actions from black hole entropy'',
arXiv:1001.4364.

\bibitem{Jacobson} T. Jacobson,
``Thermodynamics of spacetime: the Einstein equation of state'',
Phys. Rev. Lett. {\bf 75} (1995) 1260, 
gr-qc/9504004.

\bibitem{MTW} C.W. Misner, K.S. Thorne and J.A. Wheeler,
{\it Gravitation} 
(W.H. Freeman and Company, New York, 1973).

\bibitem{MarSmolin} F. Markopoulou and L. Smolin,
``Holography in a quantum spacetime'',
hep-th/9910146.

\end{thebibliography}
\end{document}